\documentclass[aps, twocolumn, showpacs, letterpaper]{revtex4}

\pdfoutput=1

\usepackage{amsmath}
\usepackage{amssymb}
\usepackage{graphicx}
\usepackage{xspace}
\usepackage[mathscr]{euscript}

\DeclareSymbolFont{SFMathGreek}{OT1}{cmss}{m}{sl} 
\DeclareMathSymbol{\slPi}{\mathalpha}{SFMathGreek}{"05} 

\newcommand{\eg}{{e.g.,\/}\xspace}
\newcommand{\ie}{{i.e.,\/}\xspace}

\newcommand{\eq}[1]{(\ref{#1})}
\newcommand{\Eq}[1]{Eq.~(\ref{#1})}
\newcommand{\Eqs}[1]{Eqs.~(\ref{#1})} 
\newcommand{\Eqsc}[2]{Eqs.~(\ref{#1}), (\ref{#2})}

\newcommand{\Sec}[1]{Sec.~\ref{#1}}

\newcommand{\App}[1]{Appendix~\ref{#1}}
\newcommand{\Fig}[1]{Fig.~\ref{#1}} 
\newcommand{\Figsc}[2]{Figs.~\ref{#1},~\ref{#2}}
\newcommand{\Ref}[1]{Ref.~\cite{#1}}
\newcommand{\Refs}[1]{Refs.~\cite{#1}}

\newcommand{\mc}[1]{\mathcal{#1}}
\newcommand{\mcc}[1]{\mathfrak{#1}}
\newcommand{\msc}[1]{\mathscr{#1}}

\newcommand{\kpt}[1]{{\kern #1 pt}}

\newcommand{\avr}[1]{\left\langle #1 \right\rangle}
\renewcommand{\vec}[1]{{\boldsymbol{\rm #1}}}

\newcommand{\pd}{\partial}
\newcommand{\const}{\text{const}}

\newcommand{\acubeeffect}{$a^{3}$-\kpt{.4}effect\xspace}
\newcommand{\supi}[1]{{{\scriptscriptstyle #1}}}
\newcommand{\parallelind}{|\kpt{-1}|}
\newcommand{\subpar}{\supi{\parallelind}}
\newcommand{\suba}[1]{_{#1\kpt{.3}}}
\newcommand{\sq}{^{\kpt{.7}2}}

\newcommand{\vg}{v\suba{g}}         % group velocity
\newcommand{\vp}{v\suba{p}}         % phase velocity
\newcommand{\Lpar}{L_\subpar}       % pulse length
\newcommand{\padiab}{\epsilon}      % small parameter
\newcommand{\Hbas}{H}               % basic Hamiltonian
\newcommand{\Pcan}{P}               % canonical momentum
\newcommand{\ptime}{\tau}           % proper time
\newcommand{\Hext}{\mcc{H}}         % extended Hamiltonian
\newcommand{\norm}[1]{\bar{#1}}     % normalization
\newcommand{\Pn}{\msc{P}}           % normalized transverse canonical momentum (aux)
\newcommand{\Pnperp}{\Pn_\perp}     % normalized transverse canonical momentum
\newcommand{\Pny}{\Pn_y}            % normalized transverse canonical momentum - y
\newcommand{\Pnz}{\Pn_z}            % normalized transverse canonical momentum - y
\newcommand{\PAng}{\theta}          % conjugate angle
\newcommand{\npx}{\rho}             % normalized kinetic momentum - x
\newcommand{\Hn}{h}                 % normalized extended Hamiltonian
\newcommand{\ampl}{a}               % normalized field
\newcommand{\ScatAng}{\chi}         % scattering angle 
\newcommand{\nvg}{\beta\suba{g}}    % normalized group velocity
\newcommand{\nvp}{\beta\suba{p}}    % normalized phase velocity
\newcommand{\Fpl}{\mc{F}}           % generating function for plasma
\newcommand{\Ppl}{\mc{P}}           % first canonical momentum in plasma
\newcommand{\Wpl}{\mc{W}}           % second canonical momentum in plasma
\newcommand{\WAng}{\psi}            % conjugate angle
\newcommand{\Hpl}{\mc{H}}           % Hamiltonian in plasma
\newcommand{\Ftr}{F}                % generating function for plasma (transformed)
\newcommand{\Sterm}{S}              % function S
\newcommand{\Ptr}{\kpt{-.5}\slPi}   % first canonical momentum in plasma (transformed)
\newcommand{\PtrAng}{\Theta}        % conjugate angle
\newcommand{\Wtr}{W}                % second canonical momentum in plasma (transformed)
\newcommand{\WtrAng}{\Psi}          % conjugate angle
\newcommand{\Htr}{\mathrm{H}}       % Hamiltonian in plasma (transformed)
\newcommand{\nPtr}{\Pi}             % \Pi/\sqrt{1-\alpha}
\newcommand{\Ff}{\msc{F}}           % generating function for the OC variables
\newcommand{\Xf}{X}                 % OC x
\newcommand{\Tf}{T}                 % OC time
\newcommand{\Pf}{\msc{P}}           % OC momentum
\newcommand{\Hf}{\msc{H}}           % OC Hamiltonian
\newcommand{\meffp}{M_*}            % effective mass in plasma
\newcommand{\meffv}{M}              % effective mass in vacuum
\newcommand{\Q}{q}                  % normalized transverse canonical momentum
\newcommand{\amax}{\ampl_{\rm max}} % a_max
\newcommand{\psim}{\norm{p}_\sim}   % normalized quiver momentum
\newcommand{\gzero}{\gamma_0}       % initial energy
\newcommand{\ginfty}{\gamma_\infty} % final energy
\newcommand{\wpl}{w}                % w=\gamma-\beta_g\rho
\newcommand{\Lperp}{L_\perp}        % pulse width
\newcommand{\Fvac}{\digamma}        % generating function for vacuum
\newcommand{\Pvac}{\rho'}           % transformed canonical momentum for vacuum
\newcommand{\wplAng}{\eta}          % canonical phase for vacuum
\newcommand{\Hvac}{h'}              % transformed Hamiltonian for vacuum

\begin{document}

\title{Ponderomotive acceleration of hot electrons in tenuous plasmas}

\author{V.~I. Geyko and G.~M. Fraiman}
\affiliation{Institute of Applied Physics RAS, 46 Ulyanov St., Nizhny Novgorod, 603950, Russia}
\author{I.~Y. Dodin and N.~J. Fisch}
\affiliation{Department of Astrophysical Sciences, Princeton University, Princeton, New Jersey 08544, USA}
\date{\today}

\begin{abstract}
The oscillation-center Hamiltonian is derived for a relativistic electron injected with an arbitrary momentum in a linearly polarized laser pulse propagating in tenuous plasma, assuming that the pulse length is smaller than the plasma wavelength. For hot electrons generated at collisions with ions under intense laser drive, multiple regimes of ponderomotive acceleration are identified and the laser dispersion is shown to affect the process at plasma densities down to $10^{17}\,\text{cm}^{-3}$. Assuming $\ampl/\gamma_g \ll 1$, which prevents net acceleration of the cold plasma, it is also shown that the normalized energy $\gamma$ of hot electrons accelerated from the initial energy $\gzero \lesssim \Gamma$ does not exceed $\Gamma \sim \ampl\gamma_g$, where $\ampl$ is the normalized laser field, and $\gamma_g$ is the group velocity Lorentz factor. Yet $\gamma \sim \Gamma$ is attained within a wide range of initial conditions; hence a cutoff in the hot electron distribution is predicted.
\end{abstract}


% 52.38.Kd -- Laser-plasma acceleration of electrons and ions
% 52.20.Fs -- Electron collisions
% 45.20.Jj -- Lagrangian and Hamiltonian mechanics 
% 41.75.Ht -- Relativistic electron and positron beams 

\maketitle

\section{Introduction} 

The recent advances in the laser technology have yielded techniques for generating electromagnetic radiation with intensities as high as $10^{22}\,\text{W}/\text{cm}^2$ \cite{ref:mourou06}. Experiments show that interaction of ultrapowerful pulses with underdense plasmas produce hot electrons with energies up to hundreds of MeV \cite{ref:mangles05, ref:chen99}. As argued in \Ref{ref:balakin06}, the effect might be due to ponderomotive acceleration of electrons, following large-angle collisions with ions in strong electromagnetic field. Assuming that the laser dispersion is negligible due to the plasma density being small, the model explains the observed power-law spectra and predicts that the particle maximum energy scales as the third power of the field amplitude. This estimate is also in approximate agreement with the available experimental data \cite{ref:balakin06}; however, the latter is insufficient to conclude whether the model is, in fact, quantitatively accurate. On the other hand, already small yet nonvanishing densities of the plasma can undermine the assumption of negligible dispersion and therefore modify the acceleration mechanism: the electron velocity can then exceed the group velocity of a laser pulse, so the particles can be reflected, or ``snow-plowed'' by the field envelope. Thus, to understand the production of hot electrons in previous and future experiments, the effect of the laser dispersion on ponderomotive acceleration must be explored.

Previously, the snow-plow acceleration  was studied in specific regimes when the electron motion becomes exactly integrable. Particularly, \Refs{ref:eloy06, ref:mendonca07, ref:mendonca09} assume equal group and phase velocities of the laser, and \Refs{ref:mckinstrie96, ref:mckinstrie97, ref:du00, ref:startsev03} suppose circular polarization and \textit{cold} electrons (\ie having zero transverse momentum), also adopted in \Refs{ref:ginzburg82, ref:ginzburg87, ref:rax92, ref:tokman99} for an oscillation-center model. However a general treatment of the relativistic ponderomotive force in plasma has not been formulated, and the effect of the laser dispersion on the ponderomotive acceleration of \textit{hot} particles has not been understood.

The focus of this paper is then twofold. First, we derive the oscillation-center (OC) Hamiltonian for a relativistic electron injected with an arbitrary momentum in a linearly polarized laser pulse propagating in tenuous plasma, assuming that the pulse length $\Lpar$ is smaller than the plasma wavelength $\lambda_p$. Second, we use this formalism to describe the ponderomotive acceleration of hot electrons generated at collisions with ions under intense laser drive. Specifically, we identify multiple regimes of this acceleration and show that the laser dispersion affects the process at plasma densities down to $n \sim 10^{17}\,\text{cm}^{-3}$. Assuming $\ampl/\gamma_g \ll 1$, which prevents net acceleration of the cold plasma, we also show that the normalized energy $\gamma$ of hot electrons accelerated from the initial energy $\gzero \lesssim \Gamma$ does not exceed $\Gamma \sim \ampl\gamma_g$, where $\ampl$ is the normalized laser field, and $\gamma_g$ is the group velocity Lorentz factor. Simultaneously, $\gamma \sim \Gamma$ is attained in a wide range of initial conditions, with the angular spread of the accelerated electrons $\chi\sim\gamma_g^{-1}$. 

Hence we conclude that the distribution of hot electrons produced at large-angle collisions with ions at $\Lpar \ll \lambda_p$ and $\ampl/\gamma_g \ll 1$ must have a cutoff at the energy $\gamma \sim \ampl\gamma_g$. This refines the result from \Ref{ref:balakin06}, showing how even weak laser dispersion can affect the acceleration gain. However, further experiments are yet needed to validate the updated scaling, because no relevant data has been reported for the regime considered here.

The paper is organized as follows. In \Sec{sec:basiceq} we introduce our basic equations. In \Sec{sec:plasma} we derive the OC Hamiltonian for a particle interacting with a laser pulse in tenuous plasma. In \Sec{sec:accel} we identify the major regimes of ponderomotive acceleration in plasma and find the general expression for the particle energy gain. In \Sec{sec:plateau}, we discuss what we call the plateau regime, where $\gamma \sim \Gamma$ is attained within a wide range of initial conditions. In \Sec{sec:conclusions} we summarize our main results. Supplementary calculations are given in Appendix.

\section{Basic equations}
\label{sec:basiceq}

Suppose a plane laser wave propagating in plasma with the group velocity $\vg$ and the phase velocity $\vp$ along the $x$ axis, so the vector potential reads $\vec{A}=\vec{y}^0\kpt{-.5}A$,
\begin{gather}
A = \mc{A}\kpt{-1}\left(\frac{x-\vg t}{\Lpar}\right)\kpt{-2.5}\cos(k[x-\vp t]).
\end{gather}
Here $\vec{y}^0$ is a unit vector along the $y$ axis, $\Lpar$ is the spatial scale of the envelope $\mc{A}$, and $k$ is the wavenumber such that $\padiab \equiv (k\Lpar)^{-1} \ll 1$. Consider a particle with mass $m$ and charge $e$ interacting with this wave, assuming that the electrostatic potential is negligible (\Sec{sec:feasib}). Then the particle Hamiltonian is~\cite{book:landau2}
\begin{gather}
\Hbas = c \sqrt{m^2c^2+p_x\sq+\left(\vec{\Pcan}_\perp -\frac{e}{c}\,\vec{A}\right)^{\!2}},
\end{gather}
where $p_x$ is the $x$ component of the particle kinetic momentum, and $\vec{\Pcan}_\perp$ is the conserved transverse canonical momentum. 

In the extended phase space, where $(t,-\Hbas)$ serves as another canonical pair and the independent variable is the proper time $\ptime$, the equivalent Hamiltonian reads \cite{my:alfven}
\begin{gather}\label{eq:extham}
\Hext = \frac{1}{2m}\,\Bigg[m^2c^2 + p_x\sq + 
\left(\Pcan_y -\frac{e}{c}\,A\right)^{\!2}+P_z^2-\frac{\Hbas^2}{c^2}\Bigg],
\end{gather}
and, numerically, ${\Hext \equiv 0}$. Introduce the dimensionless variables
\begin{subequations}
\begin{align}
\norm{x} = kx,  \qquad & \npx = p_x/mc, \\
\norm{t} = ckt, \qquad  & \gamma = \Hbas/mc^2, \\
\norm{\ptime} = kc \ptime, \qquad & \Hn = \Hext/mc^2, \\
\nvg = \vg/c, \qquad & \nvp = \vp/c,
\end{align}
\end{subequations}
and $\vec{\Pn}_\perp \equiv \vec{\Pcan}_\perp/mc = \const$. Hence we rewrite \Eq{eq:extham} as
\begin{gather}\label{eq:extham2}
\Hn = \frac{1}{2}\,\Big[1 + \npx^2 + (\Pny-\norm{A})^2 + \Pnz^2 -\gamma^2\Big],
\end{gather}
assuming $\norm{\ptime}$ is the new time, and the normalized laser field $\norm{A} \equiv eA/mc^2$ reads
\begin{gather}
\norm{A}=a\big(\padiab\kpt{.5}[\kpt{1}\norm{x}-\nvg\norm{t}\kpt{1}]\big)\cos(\norm{x}-\nvp\norm{t}).
\end{gather}

\section{Oscillation-center Hamiltonian}
\label{sec:plasma}

\subsection{Extended Hamiltonian}

Like in \Refs{ref:eloy06, ref:mendonca07, ref:mendonca09, ref:mckinstrie96, ref:mckinstrie97, ref:du00, ref:ginzburg82, ref:ginzburg87, ref:tokman99}, we assume the linear plasma dispersion, which holds for arbitrarily large~$\ampl$ at $\Lpar \ll \lambda_p$ \cite{ref:decker95, ref:sprangle90, foot:ed}. [Also, the nonlinear instabilities will be neglected as they occur on time scales exceeding the acceleration time, which is less than the wave period (\Sec{sec:hot}).] Then,
\begin{gather}\label{eq:dispersion}
\nvg=\sqrt{1-\alpha}, \quad \nvp=1/\sqrt{1-\alpha},
\end{gather}
where $\alpha = n/n_c$, and $n_c$ is the critical density~\cite{foot:waveguide}. 

Perform a canonical transformation on \Eq{eq:extham2}~\cite{foot:intact}:
\begin{gather}
(\norm{x},\npx;\,\norm{t},-\gamma)\to
(\PAng,\Ppl; \,\WAng, \Wpl),
\end{gather}
governed by the generating function
\begin{gather}
\Fpl=(\norm{x}-\nvp\norm{t})\Ppl+(\norm{x}-\nvg\norm{t})\Wpl.
\end{gather}
Then
\begin{gather}\label{eq:pgamma}
\npx = \Ppl+\Wpl, \quad \gamma=\nvp\Ppl+\nvg\Wpl,
\end{gather}
so the new Hamiltonian reads
\begin{multline}
\Hpl = \frac{1}{2}\,\Bigg\{1-\frac{\alpha\Ppl^2}{1-\alpha}+\alpha \Wpl^2+\Pnz^2+ \\
+ \big[\Pny-\ampl(\padiab\WAng)\,\cos\PAng\big]^2\Bigg\},
\end{multline}
and the new variables are given by
\begin{gather}
\PAng = \norm{x}-\nvp\norm{t}, \quad \Ppl = \frac{\gamma-\nvg\npx}{\nvp-\nvg},\label{eq:ppl}\\
\WAng = \norm{x}-\nvg\norm{t}, \quad \Wpl =- \frac{\gamma-\nvp\npx}{\nvp-\nvg}.
\end{gather}

Unlike at $\nvp=\nvg$ \cite{ref:eloy06, ref:mendonca07, ref:mendonca09}, \eg for vacuum (\App{app:vacuum}), or the exactly integrable case of circular polarization with zero $\Pny$ \cite{ref:mckinstrie96, ref:mckinstrie97, ref:du00, ref:startsev03}, there are two independent coordinates $\PAng$ and $\WAng$ entering $\Hpl$ here; hence we proceed as follows. Introduce the normalized momenta
\begin{gather}\label{eq:pplwpl}
\norm{\Ppl} = \alpha \Ppl, \quad
\norm{\Wpl} = \alpha \Wpl,
\end{gather}
which remain finite at $\alpha \to 0$; hence the Hamiltonian
\begin{multline}
\norm{\Hpl} = \frac{1}{2}\,\Bigg\{\alpha - \frac{\norm{\Ppl}^2}{1-\alpha}+\norm{\Wpl}^2+\alpha\Pnz^2+\\
+\alpha\big[\Pny-\ampl(\padiab\WAng)\,\cos\PAng\big]^2\Bigg\}.
\end{multline}
Following the general perturbation theory \cite{book:arnold, ref:dewar73, ref:johnston78, ref:johnston79}, we now seek to map out the quiver dynamics. To do this, consider a canonical transformation
\begin{gather}
(\PAng,\norm{\Ppl};\,\WAng,\norm{\Wpl})\to
(\PtrAng,\Ptr;\,\WtrAng,\Wtr)
\end{gather}
governed by the generating function
\begin{gather}
\Ftr=\PAng\Ptr+\WAng \Wtr+\Sterm(\PAng,\Ptr;\WAng,\Wtr).
\end{gather}
Choose $\Sterm$ such that $\Ptr$ and $\Wtr$ are OC canonical momenta, \ie the new Hamiltonian $\Htr(\PtrAng, \Ptr; \WtrAng, \Wtr)$ does not contain fast oscillations. Then
\begin{multline}\label{eq:S1}
\Bigg\{
-\frac{1}{1-\alpha}\Big[2\Ptr\,\pd_\PAng \Sterm+(\pd_\PAng \Sterm)^2\Big]+\\
\Big[\Wtr\,\pd_\WtrAng \Sterm+(\pd_\WtrAng \Sterm)^2\Big]^2+
\alpha f(\WAng,\PAng)\Bigg\}_\sim \! = 0,
\end{multline}
the tilde standing for the quiver part, and 
\begin{gather}
f(\WAng,\PAng) = \frac{1}{2}\,\ampl^2(\padiab\WAng)\cos 2\PAng-2\Pny\,\ampl(\padiab\WAng)\cos\PAng.
\end{gather}

At $\padiab \ll 1$, the terms containing $\pd_\WAng \Sterm$ are negligible; thus, from \Eq{eq:S1}, $\Sterm$ is nearly independent of $\Wtr$, and
\begin{gather}\label{eq:Psi}
\WtrAng = \WAng + \pd_\Wtr \Sterm \approx \WAng. 
\end{gather}
Hence \Eq{eq:S1} rewrites as
\begin{gather}\label{eq:S2}
-\frac{1}{1-\alpha}\Big[2\Ptr\,\pd_\PAng \Sterm+(\pd_\PAng \Sterm)^2-C^2\Big]+\alpha f(\WAng,\PAng)= 0,
\end{gather}
where $C^2=\avr{(\pd_\PAng \Sterm)^2}$, and the angular brackets denote averaging over $\PAng$. Solving \Eq{eq:S2} yields
\begin{gather}\label{eq:S3}
\Sterm = - \PAng\Ptr + \int^\PAng \sqrt{\Ptr^2+C^2+\alpha(1-\alpha)f(\WAng,\tilde{\PAng})}\,\,d\tilde{\PAng},
\end{gather}
where we chose the root which corresponds to $\norm{\Ppl}>0$,
\begin{gather}\label{eq:nppl}
\norm{\Ppl} = \Ptr + \pd_\PAng \Sterm.
\end{gather}
Require that $\Sterm$ does not contain a zeroth-order harmonic in $\PAng$; hence, due to \Eqs{eq:Psi}, \eq{eq:S3}, \eq{eq:nppl}, $C$ is found from
\begin{gather}\label{eq:sqrt}
\int^{2\pi}_{0} \sqrt{\Ptr^2+C^2+\alpha(1-\alpha)f(\WtrAng,\tilde{\PAng})}\,\,d\tilde{\PAng} = 2\pi \Ptr.
\end{gather}
(For an approximate solution see \Sec{sec:appr}; also see \Refs{ref:ginzburg82, ref:ginzburg87, ref:tokman99} for the case $\Pnperp=0$.) Then
\begin{gather}\label{eq:h}
\Htr = \frac{1}{2}\,\Bigg[\alpha\Big(1+\Pnperp^2+2\Phi\kpt{.3}\Big)
+\Wtr^2-\frac{\Ptr^2}{1-\alpha}\Bigg],\\
\Phi = \frac{\ampl^2}{4}\left(1-\delta^2\right),\label{eq:phi}\\
\delta^2 = \frac{2C^2}{\alpha(1-\alpha)\ampl^2}.\label{eq:delta0}
\end{gather}
Hence we integrate the motion in the variables~${(\PtrAng,\Ptr)}$:
\begin{gather}
\Ptr = \const, \quad \PtrAng = \Ptr\norm{\tau}/(1-\alpha) + \const,
\end{gather}
and the remaining canonical equations read
\begin{gather}\label{eq:wtrapprox}
\dot{\Wtr} = -\partial_\WtrAng \Htr, \quad \dot{\WtrAng} = \Wtr = \norm{\Wpl} - \pd_\WAng \Sterm \approx \norm{\Wpl}.
\end{gather}

\subsection{Effective mass $\boldsymbol{\meffp}$}

One can also revert to the space and time coordinates, which is done as follows. Apply the variable change
\begin{gather}
\Ptr = \alpha \norm{\Ptr}, \quad \Wtr = \alpha \norm{\Wtr}, \quad \Htr = \alpha \norm{\Htr},
\end{gather}
where $\norm{\Htr}(\PtrAng, \norm{\Ptr};\,\WtrAng,\norm{\Wtr})$ is the new Hamiltonian. Perform a canonical transformation
\begin{gather}
(\PtrAng, \norm{\Ptr};\,\WtrAng,\norm{\Wtr}) \to ({\Xf},{\Pf}_x;\,{\Tf},-{\Hf})
\end{gather}
governed by the generating function
\begin{gather}
\Ff =\frac{\nvp\WtrAng-\nvg\PtrAng}{\nvp-\nvg}\, {\Pf}_x-\frac{\WtrAng-\PtrAng}{\nvp-\nvg}\, {\Hf}.
\end{gather}
Then ${\Pf}_x = \avr{\npx}$, ${\Hf} = \avr{\gamma}$, and
\begin{gather}
\Xf = \frac{\nvp\WtrAng-\nvg\PtrAng}{\nvp-\nvg}, \quad
\Tf = \frac{\WtrAng-\PtrAng}{\nvp-\nvg}.
\end{gather}

Now return from the extended phase space to the physical phase space, so that $\Tf$ becomes the independent variable. Hence the new Hamiltonian
\begin{gather}\label{eq:effmassham}
\Hf=\sqrt{\meffp^2+\Pf^2}
\end{gather}
is equivalent to that of a particle with an effective mass 
\begin{gather}\label{eq:effmassp}
\meffp =  \sqrt{1+2\Phi},
\end{gather}
where $\Phi=\Phi(\ampl,\alpha,\Pny,\nPtr)$, 
\begin{gather}\label{eq:nptrdef}
\nPtr=\frac{\Ptr}{\sqrt{1-\alpha}} = \Hf-\nvg\Pf_x
\end{gather}
is a constant determined by the initial conditions, and $\Pf^2=\Pf^2_x+\Pnperp^2$ is the OC total momentum squared. Thus the average force on a particle due to the laser field, or the so-called ponderomotive force, reads
\begin{gather}\label{eq:pondforce}
\vec{F} = - \avr{\gamma}^{-1}\nabla \meffp,
\end{gather}
in the nonrelativistic case yielding $\vec{F}\approx-\nabla \Phi$, where $\Phi$ is called the ponderomotive potential \cite{my:mneg, arX:nlinphi, ref:gaponov58, ref:motz67, ref:cary77}.

From \Eq{eq:phi}, it flows that the plasmon inertia decreases the electron effective mass and $\Phi$. As the treatment is expanded to arbitrary dispersion (other polarizations are allowed, too), it can also be shown that $\meffp < \meffv$ at $\nvp>1$, and $\meffp > \meffv$ at $\nvp<1$ in the general case. However, the sign of the square root in \Eq{eq:S3} (and further) must be chosen appropriately, accounting for the fact that $\Ppl$ [\Eq{eq:ppl}] might then become negative.

\subsection{Explicit approximation for $\boldsymbol{\meffp}$}
\label{sec:appr}

\begin{figure}
\centering
\includegraphics[width=.4 \textwidth]{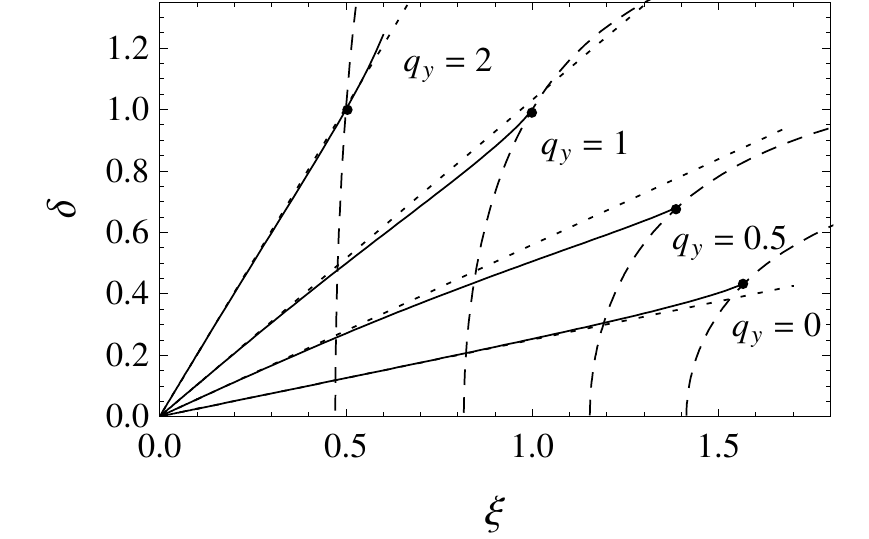}
\caption{$\delta$ [\Eq{eq:delta0}] vs. $\xi \equiv \ampl\sqrt{\alpha}/\nPtr$ for different ${\Q_y\equiv \Pny/\ampl}$: solid -- numerical solution of \Eq{eq:sqrt} for ${\xi \le \xi_*(\Q_y)}$; dotted -- analytical approximation~\eq{eq:delta}. Intersections with $\delta^2(\xi_r,\Q_\perp)$ [\Eq{eq:xistar}] (dashed) yield the reflection points $\xi_r$ (dots), and physically realizable are $\xi\le\xi_r$. Here $\Pnz = 0$ and $\alpha \to 0$; hence $\xi_r \approx \xi_*$ for $\Q_y = 0, \, 0.5, \, 1$, but $\xi_r < \xi_*$ for $\Q_y = 2$. The exact $\xi_r$ are close to those flowing from the analytical approximation, except at $\Q_y = 0.5$ here, in which case the dotted and the dashed lines do not intersect. 
}
\label{fig:cp}
\end{figure}

To find the Hamiltonian $\Htr$ and the effective mass explicitly, solve for $\delta$ using \Eq{eq:sqrt}, which rewrites as
\begin{gather}\label{eq:xistar0}
\int^{2\pi}_0 \sqrt{\xi^{-2}+\delta^2(\xi,\Q_y)/2 + 
\norm{f}(\tilde{\PAng}, \Q_y)}\kpt{2}d\tilde{\PAng} = 2\pi\xi^{-1},
\end{gather}
where
\begin{gather}\label{eq:xidef}
\xi = \ampl\sqrt{\alpha}/\nPtr,  \quad \vec{\Q} = \vec{\Pn}/\ampl, \quad \norm{f} = f/\ampl^2.
\end{gather}
At $\xi \ll 1$, $\xi\Q_y \ll 1$, this yields an approximate solution
\begin{gather}\label{eq:delta}
\delta \approx \xi\sqrt{\Q_y^2+1/16}.
\end{gather}
Then $\delta \ll 1$, so the effective mass reads
\begin{gather}\label{eq:explmeff}
\meffp = \meffv \Bigg(1-\frac{\ampl^2\delta^2}{4\meffv^2}\Bigg),
\end{gather}
where $\meffv = \sqrt{1+\ampl^2/2}$ is the effective mass in vacuum \cite{ref:kibble66, ref:tokman99, ref:bauer95, ref:bourdier01b, ref:quesnel98, my:meff, my:mneg}. Particularly, for cold particles with $\Pf=0$ (\ie $\nPtr \approx \meffv$), and $\alpha = \alpha_0/\meffv$ (assuming relativistic modification of the critical density, with $\alpha_0=\const$), one gets $\meffp = \meffv [1-\alpha_0\ampl^4/(64\meffv^5)]$, in agreement with~\Ref{ref:rax92b}.

\Eq{eq:delta} can also be extrapolated as follows. \Eq{eq:xistar0} must hold for any initial conditions; however, at large $\xi$, its right-hand side goes to zero, and on the left-hand side $\xi^{-2}$ becomes negligible in comparison with $|\norm{f}| \gtrsim 1$. On the other hand, the square root in \Eq{eq:xistar0} is supposed to remain positive and nonvanishing due to the oscillating~$\norm{f}$. Thus there is no solution for $\delta$ at $\xi \gtrsim 1$, meaning that there exists $\xi_*(\Q_y)$ such that any realizable $\xi$ satisfies
\begin{gather}\label{eq:domain}
\xi\le\xi_* \lesssim 1 
\end{gather}
(see also \Sec{sec:reflect}). Yet the exact numerical solution of \Eq{eq:xistar0} for $\delta(\xi,\Q_y)$ and its domain is close to \Eq{eq:delta} for any $\xi$ from the interval $[0,\xi_*(\Q_y)]$, as seen in \Fig{fig:cp}. Therefore \Eq{eq:delta} roughly holds for any realizable~$\xi$, and \Eq{eq:effmassp} can be used to, at least, estimate $\meffp$ explicitly.

\subsection{Reflection point}
\label{sec:reflect}

Since $\xi\le \xi_*$, a particle cannot enter a field with $\ampl > a_* \equiv \xi_* \nPtr/\sqrt{\alpha}$; thus, if the maximum field exceeds $a_*$, a particle is reflected. On the other hand,  not all $\xi$ satisfying \Eq{eq:domain} can be physically realized; thus a particle may bounce off even weaker field.

Specifically, the reflection condition is found from
\begin{gather}\label{eq:qeq}
\dot{\WtrAng}^2+ \alpha \Bigg[1+\Pnperp^2+
\frac{\ampl^2(\padiab\WtrAng)}{2}\,(1-\delta^2)\Bigg]  = \nPtr^2,
\end{gather}
which is obtained using \Eq{eq:h}, together with $\Htr \equiv 0$. Suppose $\alpha \ll \ampl^2$; then, at $\dot{\WtrAng}=0$, being the condition of particle stopping in the frame traveling with the laser envelope, \Eq{eq:qeq} yields
\begin{gather}\label{eq:xistar}
\delta^2(\xi_r,\Q_\perp) = 1+2\left(\Q^2_\perp-\xi^{-2}_r\right)
\end{gather}
for the reflection point $\xi_r$. Unlike $\xi_*(\Q_y)$, the value of $\xi_r$ is then determined by both $\Q_y$ and $\Q_z$; hence $\xi_r < \xi_*$, except at $\Q_z = 0$ and $|\Q_y| \le 1$, for which case one can show $\xi_r \to \xi_*$ for $\alpha \to 0$ (\Fig{fig:cp}). 

With \Eq{eq:delta} used as an estimate for $\delta$ \cite{foot:xi}, one can further show that, in agreement with \Refs{ref:du00, ref:mckinstrie96, ref:mckinstrie97, ref:mendonca07, ref:mendonca09},
\begin{gather}\label{eq:reflcond}
\xi_r \sim \min\{1,\, \Q_\perp^{-1}\},
\end{gather}
assuming the inequality~\eq{eq:domain}. Thus reflection is impossible at $\xi \ll \xi_r$ and possible at $\xi \sim \xi_r$, whereas larger $\xi$ cannot be realized. Therefore
\begin{gather}\label{eq:reflcond2}
\xi \le \xi_r \lesssim 1,
\end{gather}
which also yields, from \Eq{eq:delta} and $\Q_\perp \ge \Q_y$, that
\begin{gather}\label{eq:deltaineq}
\delta \lesssim 1.
\end{gather}

\section{Ponderomotive acceleration}
\label{sec:accel}

\subsection{Basic equations}

The particle energy $\gamma$, as affected by the ponderomotive force~\eq{eq:pondforce}, can now be calculated as follows. Use \Eq{eq:pgamma} together with \Eqs{eq:pplwpl} for $\Ppl$ and~$\Wpl$. Further, substitute $\norm{\Ppl}$ from \Eq{eq:nppl}, with $\Sterm$ found from \Eq{eq:S3}, and employ \Eq{eq:wtrapprox} for $\norm{\Wpl}$, with $\dot{\WtrAng}$ from \Eq{eq:qeq}; hence
\begin{widetext}
\begin{gather}\label{eq:ginfty05}
\gamma = \frac{1}{\alpha}\Big\{
\sqrt{\nPtr^2+\alpha \ampl^2\delta^2\!/2+
\alpha \big[(\ampl^2/2)\cos 2\PAng-2\Pny\,\ampl\cos\PAng\big]} \pm 
 \sqrt{\big(1-\alpha\big)\big(\nPtr^2-
 \alpha\big[1+\Pnperp^2+(\ampl^2/2)(1-\delta^2)\big]\big)}
 \kpt{1}\Big\}.
\end{gather}
\end{widetext}
Thus the energy retained outside the field is
\begin{gather}\label{eq:ginfty0}
\ginfty = \frac{1}{\alpha}
\Big\{\nPtr \pm \sqrt{\big(1-\alpha\big)\big(\nPtr^2-\alpha\big[1+\Pnperp^2\big]\big)}\kpt{1}\Big\},
\end{gather}
where the plus and the minus correspond to the particle overtaking the pulse and falling behind it, respectively. 

If no reflection occurs and the average-force approximation [from which \Eqsc{eq:ginfty05}{eq:ginfty0} are derived] holds on the time interval $(-\infty,+\infty)$, then $\ginfty$ matches the energy before entering the field, due to the conservation of $\nPtr$ and $\Pnperp$. Yet in the general case
\begin{gather}\label{eq:ginfty1}
\ginfty \sim \nPtr/\alpha,
\end{gather}
unless $\nPtr^2\gg\alpha (1+\Pnperp^2)$ \textit{and} the particle is transmitted; otherwise \Eq{eq:ginfty0} is Taylor-expanded as
\begin{gather}\label{eq:ginfty2}
\ginfty \approx \frac{1+\nPtr^2+\Pnperp^2}{2\nPtr}
\end{gather}
[cf. the exact solution \eq{eq:vgamma} for vacuum].

Hence $\ginfty$ can be found by substituting $\nPtr$ from
\begin{gather}\label{eq:nptr2}
\nPtr^2 = \wpl^2-\alpha \ampl^2\kpt{.5}\big[\norm{f}+\delta^2_0\!/2\big].
\end{gather} 
Here we employed \Eqs{eq:nppl}, \eq{eq:delta0}, \eq{eq:nptrdef}, \eq{eq:xidef} and, using \Eqsc{eq:pgamma}{eq:pplwpl},  substituted $\norm{\Ppl} = \wpl\sqrt{1-\alpha}$, with
\begin{gather}\label{eq:wpl}
\wpl= \gzero - \beta_g \npx_0
\end{gather} 
found from initial conditions (hence the index $0$). If a particle is born inside the field (\Sec{sec:hot}), the initial $\delta$ itself depends on $\nPtr$ and must be found from \Eqsc{eq:delta0}{eq:sqrt} or, approximately, from \Eq{eq:delta}; yet an estimate can be obtained as follows. From \Eqsc{eq:nptrdef}{eq:xidef}, one gets that ${\nPtr^2\!/(\alpha \ampl^2) \sim \xi^{-2} \gtrsim 1}$, the inequality being due to \Eq{eq:domain}. Together with \Eq{eq:deltaineq}, this means that, for an estimate, the term proportional to $\delta^2$ can be omitted in \Eq{eq:nptr2}, and, since $|f| \sim \max\{1,\,\Q_y\}$, one finally gets
\begin{gather}\label{eq:ptrmax}
\nPtr^2 \sim \max\{\wpl^2,\,\alpha \ampl^2,\, \alpha \ampl\Pny\}. 
\end{gather}

\subsection{Regimes of hot electron acceleration}
\label{sec:hot}

\begin{figure}
\centering
\includegraphics[width=.45 \textwidth]{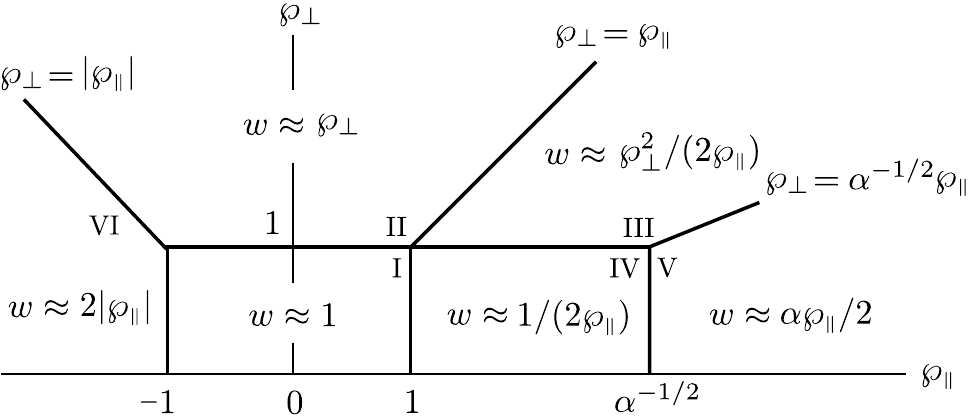}
\caption{Scalings for $\wpl$ [\Eq{eq:wpl}] depending on the normalized initial momentum $\vec{\wp}$ ($\wp_z=0$, ${\alpha \ll 1}$). The roman numbers tag distinct regimes.}
\label{fig:w}
\end{figure}

Consider a hot electron produced inside a laser pulse, \eg due to ionization or collision (\Sec{sec:vacuum}), at some~$\PAng_0$ and~$\ampl$ of the order of the maximum amplitude $\amax$. Hence, as the particle starts to oscillate, it attains ${\gamma \sim \ginfty}$ already on a fraction of the oscillation period [\Eq{eq:ginfty05}], like described in \Ref{my:gev}. To calculate the associated energy gain, suppose an initial momentum $\vec{\wp} \equiv \vec{p}_0\kpt{-.5}/mc$, for simplicity assuming $\wp_z=0$ and $\alpha \ll 1$; thus
\begin{gather}\label{eq:pperp}
\Pny = \wp_\perp - \ampl\cos\PAng_0 \sim \max\{\wp_\perp,\, \ampl\},
\end{gather} 
whereas $\wp_\subpar$ will denote the $x$-component of the particle kinetic momentum. Then one of the six regimes is realized, depending on how $\wpl$ [\Eq{eq:wpl}] is expanded~(\Fig{fig:w}), and more regimes appear due to \Eqsc{eq:ptrmax}{eq:pperp} allowing multiple scalings for $\nPtr$ and $\Pny$, respectively. 

\begin{figure*}
\begin{center}
\includegraphics[width=.95 \textwidth]{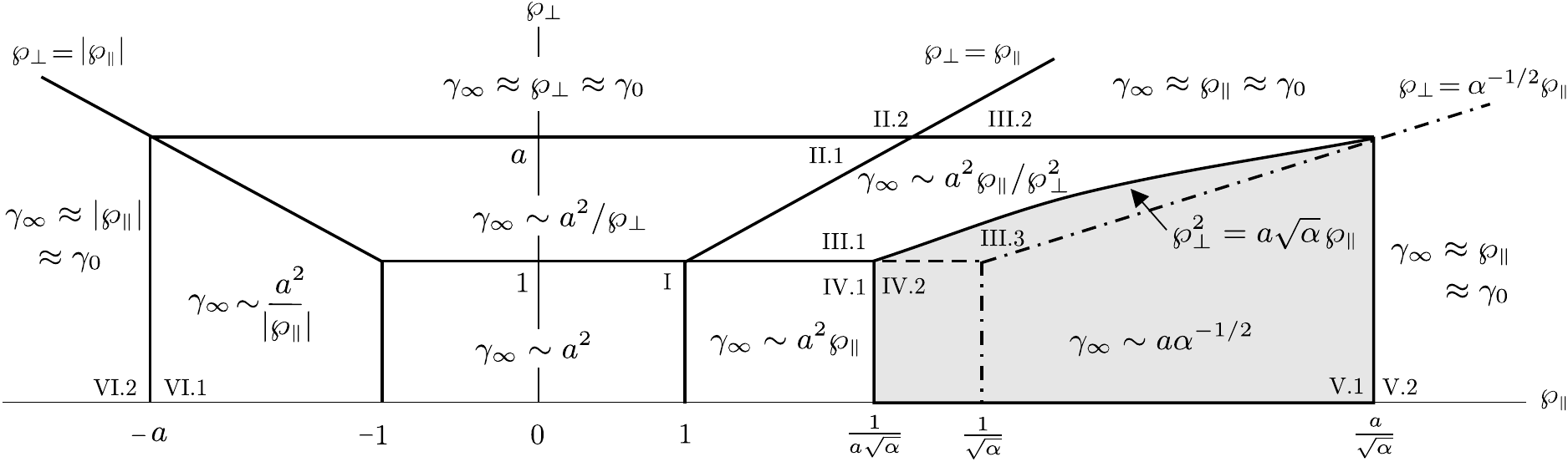}
\\[20pt]
\begin{tabular*}{.9\textwidth}{
 @{\extracolsep{\fill}} c 
 @{\extracolsep{\fill}} c 
 @{\extracolsep{\fill}} c 
 @{\extracolsep{\fill}} c 
 @{\extracolsep{\fill}} c
 @{\extracolsep{\fill}}
 } 
\hline \hline
\kpt{5} Regime & 
$\wpl$ [\Eq{eq:wpl}] & 
$\Pny$ [\Eq{eq:pperp}]&  
$\nPtr$ [\Eq{eq:nptr2}]& 
$\ginfty$ [\Eq{eq:ginfty0}]\quad
\\
\hline
I &
$\wpl \approx 1$ &
$\Pny \sim \ampl$ &
$\nPtr \approx 1$ &
$\ginfty \sim \ampl^2$
\\
II.1 &
$\wpl \approx \wp_\perp$ &
$\Pny \sim \ampl$ &
$\nPtr \approx \wp_\perp$ &
$\ginfty \sim \ampl^2/\wp_\perp$
\\
II.2 &
$\wpl \approx \wp_\perp$ &
$\Pny \approx \wp_\perp$ &
$\nPtr \approx \wp_\perp$ &
$\ginfty \approx \wp_\perp \approx \gzero$
\\
III.1 &
$\wpl \approx \wp_\perp^2/(2\wp_\subpar)$ &
$\Pny \sim \ampl$ &
$\nPtr \approx \wp_\perp^2/(2\wp_\subpar)$ &
$\ginfty \sim \ampl^2\wp_\subpar/\wp_\perp^2$
\\
III.2 &
$\wpl \approx \wp_\perp^2/(2\wp_\subpar)$ &
$\Pny \approx \wp_\perp$ &
$\nPtr \approx \wp_\perp^2/(2\wp_\subpar)$ &
$\ginfty \approx \wp_\subpar \approx \gzero$
\\
III.3 &
$\wpl \approx \wp_\perp^2/(2\wp_\subpar)$ &
$\Pny \sim \ampl$ &
$\nPtr \sim \ampl \alpha^{1/2}$ &
$\ginfty \sim \ampl\alpha^{-1/2}$
\\
IV.1 &
$\wpl \approx 1/(2\wp_\subpar)$ &
$\Pny \sim \ampl$ &
$\nPtr \approx 1/(2\wp_\subpar)$ &
$\ginfty \sim \ampl^2\wp_\subpar$
\\
IV.2 &
$\wpl \approx 1/(2\wp_\subpar)$ &
$\Pny \sim \ampl$ &
$\nPtr \approx 1/(2\wp_\subpar)$ &
$\ginfty \sim \ampl\alpha^{-1/2}$
\\
V.1 &
$\wpl \approx \alpha\wp_\subpar/2$ &
$\Pny \sim \ampl$ &
$\nPtr \sim \ampl \alpha^{1/2}$ &
$\ginfty \sim \ampl\alpha^{-1/2}$
\\
V.2 &
$\wpl \approx \alpha\wp_\subpar/2$ &
$\Pny \sim \max\{\wp_\perp,\, \ampl\}$ &
$\nPtr \approx \alpha\wp_\subpar/2$&
$\ginfty \approx \wp_\subpar \approx \gzero$
\\
VI.1 &
$\wpl \approx 2|\wp_\subpar|$ &
$\Pny \sim \ampl$ &
$\nPtr \approx 2|\wp_\subpar|$&
$\ginfty \sim \ampl^2/|\wp_\subpar|$
\\
VI.2 &
$\wpl \approx 2|\wp_\subpar|$ &
$\Pny \sim \max\{\wp_\perp,\, \ampl\}$ &
$\nPtr \approx 2|\wp_\subpar|$&
$\ginfty \approx |\wp_\subpar| \approx \gzero$
\\
\hline \hline
\end{tabular*}
\end{center}
\caption{Regimes of ponderomotive acceleration for electrons born inside laser field with initial momentum $\vec{\wp}$. The dashed and dot-dashed lines separate different domains corresponding to the same energy~$\ginfty$. The dot-dashed line is also a schematic of the curve~\eq{eq:reson}, at which the particle velocity equals the pulse group velocity. Shaded is the ``plateau'' where the maximum energy $\ginfty \sim a\alpha^{-1/2}$ is independent of $\vec{\wp}$. The roman numbers are the same as for the corresponding domains in \Fig{fig:w}.}
\label{fig:regimes}
\end{figure*}

Below we limit our consideration to only a part of these regimes, because of the following. According to \Eqsc{eq:reflcond}{eq:reflcond2}, a pulse with a maximum amplitude satisfying $\alpha\amax^2 \gtrsim 1$ will snow-plow cold electrons of the background plasma, which have ${\nPtr = 1}$ and ${\Pnperp=0}$~\cite{foot:xi}. However, this would result in a significant electrostatic potential (ahead of the pulse) which is not included into the model; thus we assume 
\begin{gather}\label{eq:qn}
\alpha\amax^2 \ll 1,
\end{gather} 
so only few, hot electrons could be snow-plowed. Assuming also $\ampl \gg 1$, twelve distinct regimes persist (\Fig{fig:regimes}), and those of primary interest are discussed below. 

\subsection{Acceleration in vacuum}
\label{sec:vacuum}

Suppose that an electron is produced at rest, \eg due to ionization \cite{ref:hu02, ref:hu06}; then, from \Eq{eq:pperp},
\begin{gather}\label{eq:q0}
\Pnperp = \ampl_0 \equiv \ampl \cos\PAng_0 \sim \ampl.
\end{gather}
At $\alpha \ll 1$, the pulse travels much faster than the particle; hence the weak dispersion due to plasma is inessential in this case. Then \Eq{eq:nptr2} yields $\nPtr \approx \wpl = 1$, so $\xi \sim a\sqrt{\alpha}\ll 1$ and $\nPtr^2 \gg \alpha (1+\Pnperp^2)$, both because of \Eq{eq:qn}. Therefore particle reflection from the pulse is impossible in this case (\Sec{sec:reflect}), and \Eq{eq:ginfty2} applies, yielding
\begin{gather}\label{eq:a2}
\ginfty = 1+\ampl_0^2/2,
\end{gather}
in agreement with \Ref{my:gev} and regime I in \Fig{fig:regimes}.

When a particle is born with positive $\wp_\subpar \gg 1$, stronger acceleration is predicted from \Eq{eq:ginfty2} due to reduced~$\nPtr$. Indeed, suppose a small pitch angle $\ScatAng_0 \approx \wp_\perp/\wp_\subpar$ and, again, neglect the plasma dispersion ($\alpha \to 0$); then
\begin{gather}\label{eq:wplcc}
\wpl \approx \frac{1+\wp_\perp^2}{2\wp_\subpar} \ll 1.
\end{gather}
Similarly, \Eq{eq:ginfty2} holds, so one gets
\begin{gather}\label{eq:a31}
\ginfty \approx \frac{\Pny^2\wp_\subpar}{1+\wp_\perp^2} \sim
 \frac{a^2\wp_\subpar}{1+\ScatAng_0^2\wp^2_\subpar},
\end{gather}
covering regimes III.1 and IV.1 in \Fig{fig:regimes}. Hence only a small fraction of electrons is accelerated efficiently, particularly those with $\ScatAng_0 \lesssim \wp_\subpar^{-1} \ll 1$. However, the maximum energy now scales as $\ginfty \sim a^2\wp_\subpar$, which is bigger than that flowing from \Eq{eq:a2} by the factor $\wp_\subpar \gg 1$.

Specifically, the described effect is anticipated at large-angle electron-ion collisions in tenuous plasmas \cite{ref:balakin06}. Suppose a cold electron oscillating in a laser field with a quiver kinetic momentum $\psim \sim \ampl$ and zero average velocity. (For the general case, see \Fig{fig:regimes} and \Sec{sec:plateau}.) Suppose further that this particle collides with an ion such that the momentum vector instantaneously rotates toward the pulse propagation direction,~\ie
\begin{gather}\label{eq:cc}
\wp_\subpar \approx \psim, \quad
\wp_\perp \approx \psim \ScatAng_0.
\end{gather}
Then the maximum $\ginfty$ from \Eq{eq:a31} reads 
\begin{gather}\label{eq:a3}
\ginfty \sim \ampl^3,
\end{gather}
the result being called the \acubeeffect \cite{ref:balakin06}, and the angular spread of the accelerated electrons~is
\begin{gather}\label{eq:vscang}
\ScatAng \approx \Pnperp/ \ginfty \sim \ampl^{-2} \ll 1.
\end{gather}

\subsection{Modification of the $\boldsymbol{a}^{\boldsymbol{3}}$-effect in plasma}
\label{sec:a3pl}

Increasing the number of accelerated electrons requires higher plasma densities, and the \acubeeffect is modified in this case because of the laser dispersion; hence the energy gain is calculated differently. Particularly, for electrons with the initial conditions~\eq{eq:cc}, one has ${\wpl \sim \ampl^{-1}}$ [\Eq{eq:wplcc}; regime~IV] and ${\Q_y \sim 1}$; then \Eq{eq:ptrmax} yields $\nPtr^2 \sim \max\{\alpha \ampl^2,\, \ampl^{-2}\}$. At ${\sigma \equiv \alpha\amax^4 \ll 1}$ (regime~IV.1), one obtains $\nPtr \sim \ampl^{-1}$, so the reflection condition is not met, and the plasma effect is negligible. Suppose now that $\sigma \gtrsim 1$ (regime~IV.2). Then one gets 
\begin{gather}\label{eq:nptr}
\nPtr \sim \ampl\sqrt{\alpha},
\end{gather} 
so it becomes possible to reflect electrons from the pulse, at least, for some $\PAng_0$. (In vacuum, this effect is impossible because particles could not travel faster than light.) Hence the final energy is estimated from \Eq{eq:ginfty1}~as
\begin{gather}\label{eq:scale1}
\ginfty  \sim \ampl\alpha^{-1/2},
\end{gather}
and the angular spread of the accelerated electrons is
\begin{gather}\label{eq:pscang}
\ScatAng \approx \Pnperp/ \ginfty \sim \sqrt{\alpha} \ll 1.
\end{gather}

\begin{figure}
\centering
\includegraphics[width=.45 \textwidth]{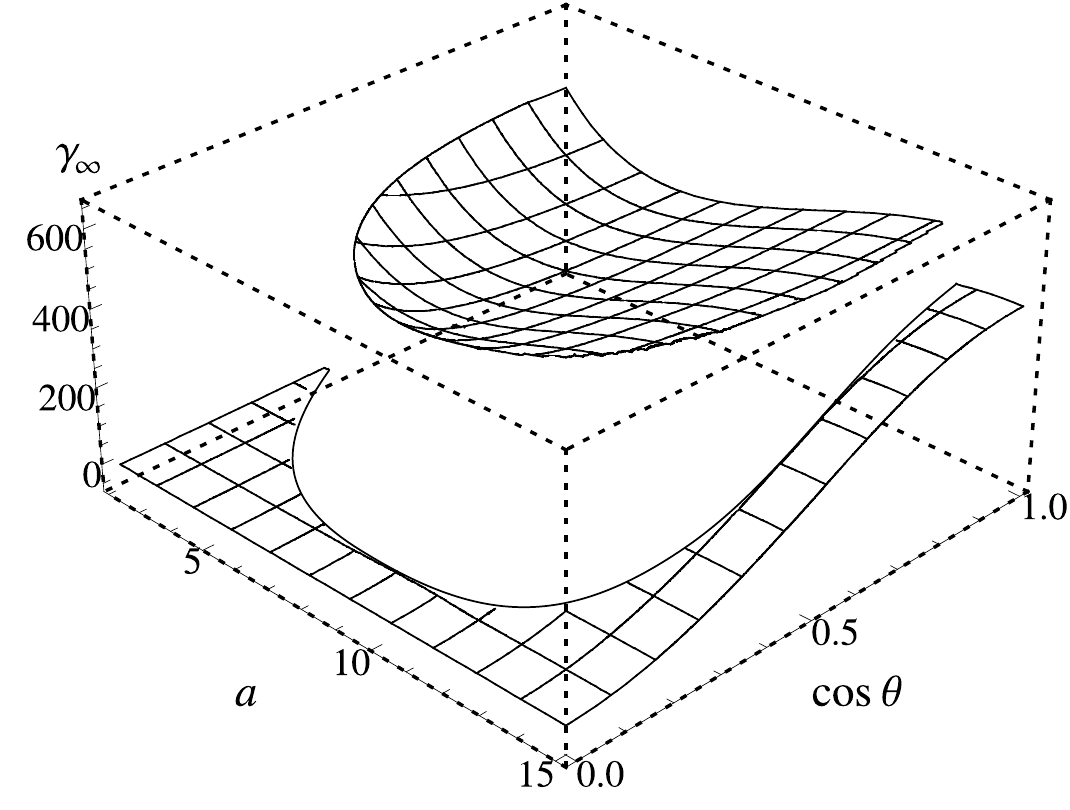}
\caption{The final energy $\ginfty = p_x/mc$ of electrons accelerated by a plane laser pulse in tenuous plasma vs. the normalized vector potential envelope $a=e\mc{A}/mc^2$ at collision and cosine of the collision phase $\PAng_0$; ${\alpha=10^{-3}}$, $a_{\rm max} = 15$. The elevation corresponds to the electrons being snow-plowed.}
\label{fig:gray}
\end{figure}

Now rewrite \Eq{eq:scale1} as $\ginfty \sim \ampl^3 \sigma^{-1/2}$. Then a uniform scaling is obtained, which covers both regimes IV.1 and IV.2, accounting for how the \acubeeffect is modified with the plasma density:
\begin{gather}\label{eq:scale}
\ginfty  \sim \ampl^3 \times \min \{1,\,\sigma^{-1/2}\}.
\end{gather}
This agrees with the results of our numerical simulations. Specifically, at $\sigma \ll 1$ we observed the vacuum \acubeeffect, and electron reflection from a pulse was seen at 
\begin{gather}
\alpha\amax^4 \gtrsim 4.4. 
\end{gather}
Hence a sharp dependence of $\ginfty$ on whether particles are reflected or not [albeit the \textit{scaling} holds for reflected and transmitted electrons equaly, as predicted from \Eqsc{eq:ginfty0}{eq:ginfty1}] and the abrupt elevation in \Figsc{fig:gray}{fig:elevation}, both agreeing with \Eqsc{eq:scale1}{eq:scale}.

\begin{figure}
\centering
\includegraphics[width=0.45 \textwidth]{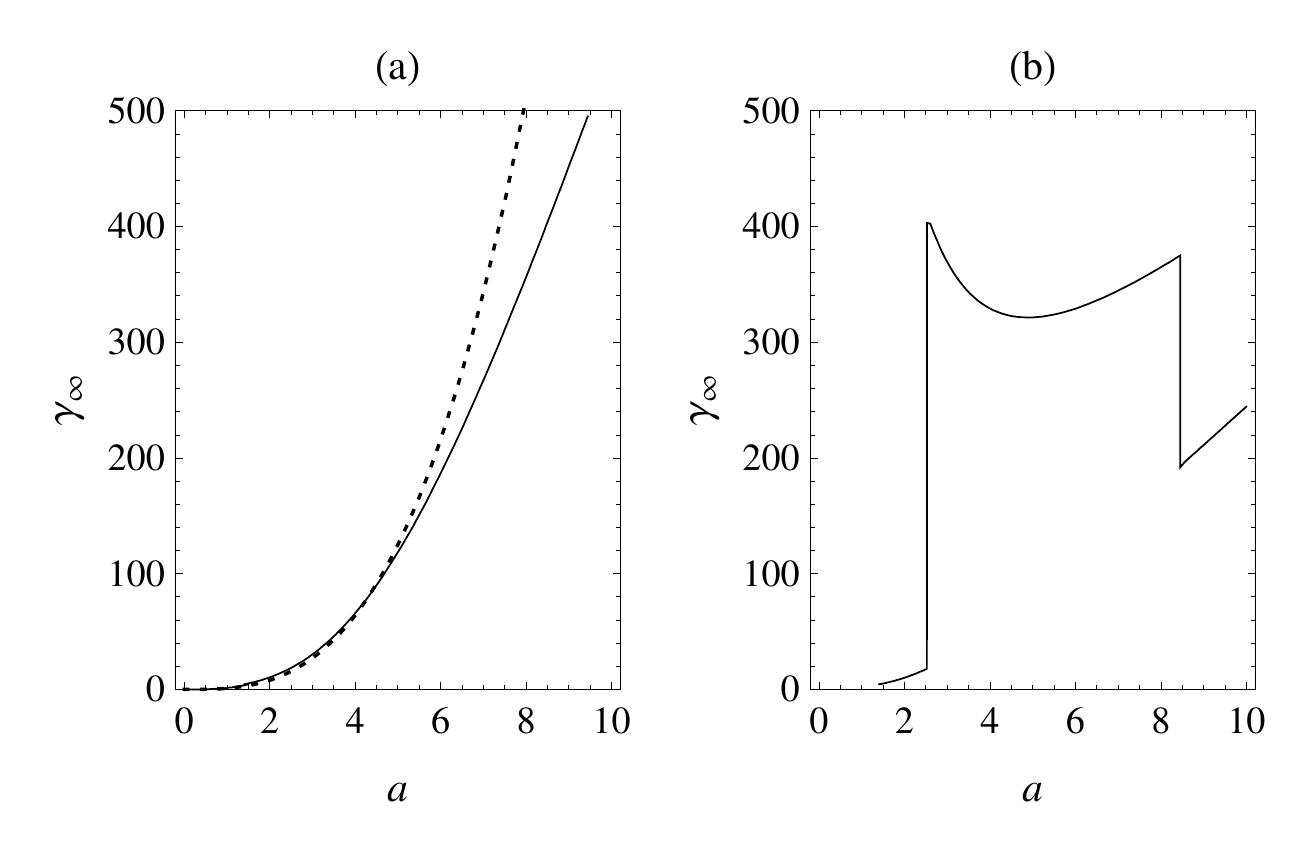}
\caption{Same as in \Fig{fig:gray} for $\PAng_0=0$, with $\amax=10$. (a)~${\alpha = 10^{-4}}$: numerical (solid) and analytical, $\ginfty = \ampl^3$, corresponding to the \acubeeffect (dashed); no particle reflection from the pulse. (b)~${\alpha = 10^{-3}}$ (numerical); the elevation corresponds to the electrons being snow-plowed.} 
\label{fig:elevation}
\end{figure}

\section{Plateau regime}
\label{sec:plateau}

\subsection{Maximum energy gain}
\label{sec:limit}

Now consider a more realistic case when the electron is also preaccelerated by the pulse \textit{before} the collision; hence we assume arbitrary initial conditions instead of \Eq{eq:cc}. Similarly to \Sec{sec:a3pl}, one can show that the acceleration is affected by plasma only in regimes~III.3, IV.2, V.I, and V.2 (\Fig{fig:regimes}). Those adjoin the curve 
\begin{gather}\label{eq:reson}
\wp_\subpar = \sqrt{\frac{1+\wp_\perp^2}{\alpha}},
\end{gather} 
which corresponds to the particle traveling at the pulse group velocity, with $\wp_\perp \ll \wp_\subpar$ (dot-dashed in \Fig{fig:regimes}). Hence the respective interactions are classified as follows.
\begin{itemize}
\item
In regimes III.3 and IV.2, a particle initially travels along $x$-axis slower than the pulse and is accelerated up to the energy~\eq{eq:scale1}.
\item
In regime V.I, a particle initially travels along $x$-axis faster than the pulse. However, it gains additional transverse momentum before it escapes from the field, resulting in the same energy gain~\eq{eq:scale1}.
\item
In regime V.2, a particle is fast enough to run ahead of the pulse such that the energy $\gamma$ is not affected ($\ginfty \approx \gzero$), as opposed to vacuum where $\ginfty \approx \gzero\ampl^2$ would apply at arbitrarily large $\gzero$ (cf.~IV.1). 
\item
In all other regimes, the particle gains energy smaller than both $\gzero$ and that given by \Eq{eq:scale1}.
\end{itemize}
Thus for an electron born inside a laser field one has
\begin{gather}\label{eq:abssc}
\ginfty  \lesssim \max\{\gzero, \, \Gamma\},
\end{gather}
where $\Gamma\sim \ampl\gamma_g$ is the energy of a particle comoving with the pulse, with the transverse momentum ${\norm{p}_\perp \sim a \gg 1}$ and the group velocity Lorentz factor ${\gamma_g = \alpha^{-1/2}}$. 

Assuming $\gzero < \Gamma$, the maximum (over $\PAng_0$) of the particle final energy is then attained in the ``plateau'' formed by the domains III.3, IV.2, V.I, where it is independent of the initial momentum $\vec{\wp}$ and so is the angular spread of the accelerated electrons:
\begin{gather}\label{eq:Gamma}
\ginfty \sim \ampl\gamma_g, \quad
\ScatAng \sim \gamma_g^{-1}.
\end{gather}
Below we assess the feasibility of the plateau regime and suggest an estimate for the energy of hot electrons which can be produced in conceivable experiments.

\subsection{Required parameters}
\label{sec:feasib}

\begin{figure}
\centering
\includegraphics[width=0.45 \textwidth]{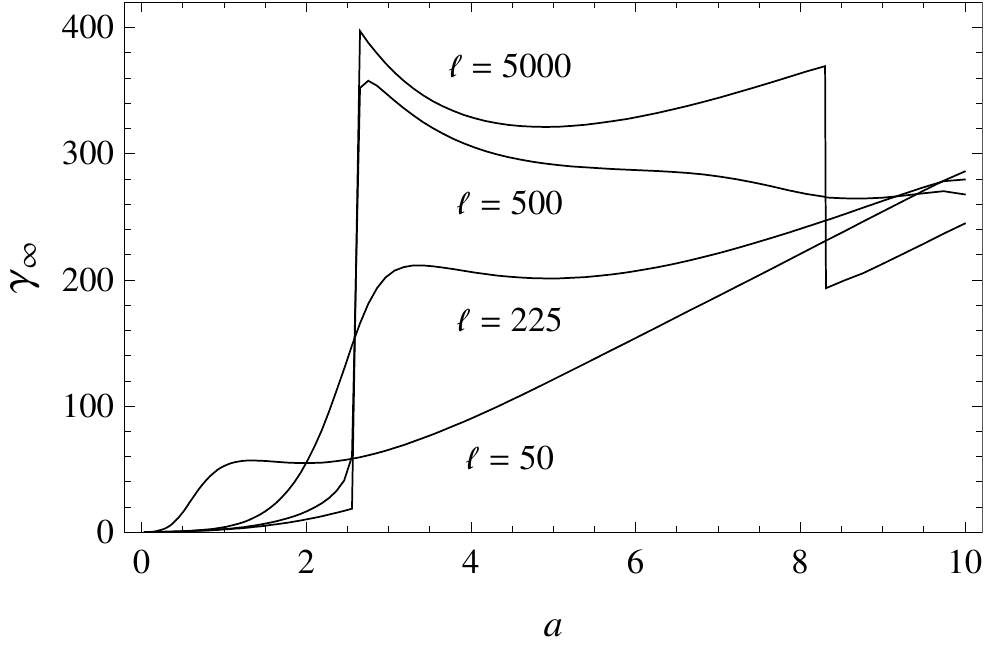}
\caption{Same as in \Fig{fig:gray} for $\PAng_0=0$ and $\amax=10$, but for a two-dimensional pulse $\ampl = \amax \exp(-x^2/\Lpar^2-y^2/\Lperp^2)$ with $\Lpar=11\lambda$ and different $\ell \equiv \Lperp/\lambda$, where $\lambda = 2\pi/k$.}
\label{fig:narrow}
\end{figure}

The one-dimensional (1D) model above neglects electron escape from the accelerating field in the transverse direction. This is a valid approximation if
\begin{gather}\label{eq:klperp}
\Delta\norm{\tau} \lesssim k\Lperp/\norm{p}_\perp,
\end{gather}
where $\Delta\norm{\tau} \sim \Delta \WtrAng/\dot{\WtrAng}$ is the normalized proper time of the interaction, and $\Delta \WtrAng \sim 1$ because the acceleration occurs on a single period (\Sec{sec:hot}). In the plateau regime, \Eq{eq:qeq} yields ${\dot{\WtrAng} \sim \ampl\sqrt{\alpha}}$; thus \Eq{eq:klperp} rewrites~as
\begin{gather}
\Lperp/\lambda \gtrsim \big(2\pi\sqrt{\alpha}\,\big)^{-1},
\end{gather}
where we took $\lambda$ for the laser wavelength, and ${\norm{p}_\perp \sim a}$. For narrower pulses, the energy gain would be somewhat lower than that predicted by \Eq{eq:Gamma}, particularly for particles born at $\norm{A} \ll \amax$, as also confirmed in our numerical simulations (\Fig{fig:narrow}). Nonetheless one can anticipate the 1D scaling to hold for feasibly focused ultraintense fields down to about $\alpha \sim 10^{-4}$. Hence the laser dispersion should affect the electron acceleration at plasma densities down to about $10^{17}\,\text{cm}^{-3}$.

Now let us estimate the influence of the previously neglected wake potential $\varphi$, which impedes the acceleration because the associated electrostatic force is directed oppositely to the ponderomotive force \cite{ref:du00, ref:esirkepov06, ref:shvets00}. The energy gain due to the electric field $\vec{E}_\varphi = -\nabla \varphi$ is $\gamma_\varphi \sim e E_\varphi L/mc^2$, where $L \sim k^{-1}\Delta\norm{\tau} \ginfty$ is the interaction length, or $kL \sim \gamma_g^2$. Assuming the wake spatial scale of about the plasma wavelength $\lambda_p = (k\sqrt{\alpha}\,)^{-1}$ and the density perturbation of the order of $n$, the Poisson's equation gives $a_\varphi \equiv eE_\varphi/(mc\omega) \sim \sqrt{\alpha}$. Then $\gamma_\varphi \sim a_\varphi \,kL \sim \alpha^{-1/2}$, yielding $\gamma_\varphi/\ginfty \sim a^{-1} \ll 1$, \ie the wake is indeed negligible~\cite{foot:intlength}.

Hence \Eq{eq:Gamma} is a valid approximation for estimating the electron final energy. For example, at laser intensity $I \sim 10^{20}\,\text{W}/\text{cm}^2$ and wavelength $\lambda \sim 1\,\mu\text{m}$, corresponding to $a \approx [\lambda/(1\,\mu\text{m})][I/(1.37\times 10^{18}\,\text{W}/\text{cm}^2)]^{1/2} \sim 10$, and $n \sim 10^{17}\,\text{cm}^{-3}$, corresponding to $\alpha \sim 10^{-4}$, \Eqs{eq:Gamma} predict $\ginfty \sim 10^3$ and $\ScatAng \sim 0.01$. Therefore hot electrons can be accelerated to energies of a fraction of GeV and will be scattered within a small angle of $0.6^{\circ}$.

\section{Conclusions}
\label{sec:conclusions}

In this paper, we derive the oscillation-center Hamiltonian for an electron injected with an arbitrary momentum in a linearly polarized laser pulse propagating in tenuous plasma, assuming that the pulse length $\Lpar$ is smaller than the plasma wavelength $\lambda_p$. We then use this formalism to describe the ponderomotive acceleration of hot electrons generated at collisions with ions under intense laser drive. Specifically, we identify multiple regimes of this acceleration and show that the laser dispersion affects the process at plasma densities down to $10^{17}\,\text{cm}^{-3}$. Assuming $\ampl/\gamma_g \ll 1$ [\Eq{eq:qn}], which prevents net acceleration of the cold plasma, we also show that the normalized energy $\gamma$ of electrons accelerated from the initial energy $\gzero \lesssim \Gamma$ does not exceed $\Gamma \sim \ampl\gamma_g$, where $\ampl$ is the normalized laser field, and $\gamma_g$ is the group velocity Lorentz factor. Simultaneously, $\gamma \sim \Gamma$ is attained in a wide range of initial conditions, with the angular spread of the accelerated electrons ${\chi\sim\gamma_g^{-1}}$. Hence the distribution of hot electrons produced at large-angle collisions with ions at $\Lpar \ll \lambda_p$ and $\ampl/\gamma_g \ll 1$ will have a cutoff at $\gamma \sim \ampl\gamma_g$. This refines the result from \Ref{ref:balakin06}, showing how even weak laser dispersion can affect the acceleration gain. However, further experiments are yet needed to validate the updated scaling, because no relevant data has been reported for the regime considered here.

%\gap

\section{Acknowledgments}

This work was supported by the Russian Foundation for Basic Research through Grant No.~08-02-01209-a and the NNSA under the SSAA Program through DOE Research Grant No.~DE-FG52-04NA00139.

%\gap

\appendix

\section{Energy gain in vacuum}
\label{app:vacuum}

In the case of vacuum, a simplified solution is possible as follows. Perform a canonical transformation \cite{foot:intact}
\begin{gather}
(\norm{x},\npx;\,\norm{t},-\gamma)\to
(\PAng,\Pvac;\,\wplAng, -\wpl)
\end{gather}
using a generating function
\begin{gather}
\Fvac = (\norm{x}-\norm{t})\Pvac - \norm{t}\kpt{.7}\wpl.
\end{gather}
Then
\begin{gather}
\PAng = \norm{x}-\norm{t}, \quad \Pvac=\npx, \quad
\wplAng = \norm{t}, \quad \wpl = \gamma-\npx,
\end{gather}
and the transformed extended Hamiltonian is given~by
\begin{gather}\label{eq:extham4}
\Hvac = 1 -\wpl^2-2\wpl\Pvac+\Pnz^2+\big\{\Pny-\ampl(\padiab\PAng)\,\cos\PAng\big\}^2 \equiv 0\nonumber.
\end{gather}
Then $\wpl$ is conserved, yielding an explicit solution for~$\npx$:
\begin{gather}\label{eq:vacP}
\npx = \frac{1}{2\wpl}\Big[1-\wpl^2+\Pnz^2+\big\{\Pny-\ampl(\padiab\PAng)\,\cos\PAng\big\}^2\Big].
\end{gather}
Hence the particle energy $\gamma = \npx + \wpl$ is obtained, and outside the field one has [cf.~\Eq{eq:ginfty2}]
\begin{gather}\label{eq:vgamma}
\ginfty = \frac{1+\wpl^2+\Pnperp^2}{2\wpl}.
\end{gather}


\begin{thebibliography}{10}
\bibitem{foot:ed}
The numerical simulations reported in \Ref{ref:startsev03} confirm that the
  linear plasma model is reasonably accurate for calculating the electron
  ponderomotive acceleration by a laser pulse with $\ampl \gg 1$ and $\Lpar \ll
  \lambda_p$.
\bibitem{foot:waveguide}
The dispersion \eq{eq:dispersion} due to plasma [\cite{book:jackson}, Sec.~7.9]
  can also be attributed to a waveguide structure [\cite{book:jackson},
  Sec.~8.3].
\bibitem{foot:intact}
The transverse variables remain intact, so the corresponding terms in the
  generating functions are omitted.
\bibitem{foot:xi}
\Eq{eq:delta} is not accurate enough to predict the actual reflection point at
  $\Q_y \sim 1$ (except at $\Q_z \gg 1$); \eg \Fig{fig:cp} shows no common
  solution for \Eq{eq:delta} and \Eq{eq:xistar} at $\Q_y = 0.5$. On the other
  hand, at $\Q_\perp = 0$, the latter equations yield $\xi^2_r =
  4(2-\sqrt{2}\,)\approx 2.34$, which agrees with the results of our
  single-particle simulations.
\bibitem{foot:intlength}
For interaction on the whole pulse length (${L \sim \gamma_g^2 \Lpar}$), one
  can similarly obtain $\gamma_\varphi/\ginfty \sim \Lpar/(\lambda\ampl)$.
  Thus, for ultrashort intense pulses, $\varphi$ is, again, negligible.
\end{thebibliography}
\end{document}